\documentstyle[11pt,aaspp4,flushrt]{article}

\newcommand{\cmjj}{\mbox{${\rm cm^{-2}}$}}

\newcommand{\kms}{\mbox{km\ s${^{-1}}$}}
\newcommand{\lya}{\mbox{${\rm Ly}\alpha$}}

\begin{document}

\title{AN EMPIRICAL LIMIT ON EXTREMELY HIGH REDSHIFT GALAXIES\altaffilmark{1,2}}

\author{KENNETH M. LANZETTA and AMOS YAHIL \\
Department of Physics and Astronomy, State University of New York at Stony
Brook \\
Stony Brook, NY 11794--3800, U.S.A.}

\and

\author{ALBERTO FERN\'{A}NDEZ-SOTO \\
School of Physics, University of New South Wales \\
P.O. Box 1, Kensington, NSW 2033, AUSTRALIA}

\altaffiltext{1}{Based on observations with the NASA/ESA Hubble Space
Telescope, obtained at the Space Telescope Science Institute, which is operated
by the Association of Universities for Research in Astronomy, Inc., under NASA
contract NAS5--26555.}

\altaffiltext{2}{Based on observations made at the Kitt Peak National
Observatory, National Optical Astronomy Observatories, which is operated by the
Association of Universities for Research in Astronomy, Inc.\ (AURA) under
cooperative agreement with the National Science Foundation.}

\newpage

\begin{abstract}

  We apply the Lyman absorption signature to search for galaxies at redshifts $z
\approx 6 - 17$ using optical and infrared images of the Hubble Deep Field.  
The infrared images are sensitive to a point source $5 \sigma$ detection
threshold of $AB(22,000) = 23.8$, which adopting plausible assumptions to relate
rest-frame ultraviolet flux densities to unobscured star formation rates is
easily sufficient to detect the star formation rates expected for massive
elliptical galaxy formation to quite high redshifts.  For $q_0 = 0.5$, the
infrared images are sensitive to an unobscured star formation rate of $\dot{M} =
100 \ h^{-2} \ M_\odot$ yr$^{-1}$ to redshifts as large as $z = 17$, and for
$q_0 = 0$, the infrared images are sensitive to an unobscured star formation
rate of $\dot{M} = 300 \ h^{-2} \ M_\odot$ yr$^{-1}$ to redshifts as large as $z
= 14$.  The primary result of the analysis is that only one extremely high
redshift galaxy candidate is identified at the $5 \sigma$ level of significance
(and four at the $4 \sigma$ level).  This implies a strict upper limit to the
surface density of extremely high redshift galaxies of $< 1.5$ arcmin$^{-2}$ to
a limiting magnitude threshold $AB(22,000) = 23.8$.  This also implies a strict
upper limit to the volume density of extremely high redshift galaxies if (and
only if) such galaxies are not highly obscured by dust.

\end{abstract}
 

\newpage

\section{INTRODUCTION}

  A standard paradigm is that massive elliptical galaxies form most or all of
their stars in a single rapid burst at early epochs.  According to this
scenario, early elliptical galaxies undergo a phase of intense ultraviolet
luminosity, as the star formation rates required to produce $\approx 10^{11}$
stars in under $\approx 10^9$ yr exceed 100 $M_\odot$ yr$^{-1}$.  Recent
progress in identifying galaxies at high redshifts has opened up the possibility
of directly observing this epoch of massive elliptical galaxy formation.  In
particular, various analyses of the Hubble Deep Field (HDF) images obtained by
the Hubble Space Telescope (HST) have identified galaxies with spectroscopically
determined redshifts as large as $z \approx 4$ (Steidel et al.\ 1996; Lowenthal
et al.\ 1997) and photometrically estimated redshifts as large as $z \approx 6$
(Lanzetta, Yahil, \& Fern\'{a}ndez-Soto 1996; Lanzetta, Fern\'{a}ndez-Soto, \&
Yahil 1997).  A primary result of these analyses is that the unobscured star
formation rates of high-redshift galaxies inferred directly from the observed
ultraviolet continua are surprisingly modest---typically less than $\approx 10$
$M_\odot$ yr$^{-1}$ and always well less than the $\approx 100$ $M_\odot$
yr$^{-1}$ or more expected for massive elliptical galaxy formation.  Various
proposals have been put forward to explain the apparent discrepancy between the
expected and observed star formation rates, including the possibilities that (1)
massive elliptical galaxies form at lower redshifts, say $z < 2.5$ (Madau et
al.\ 1996), (2) massive elliptical galaxies form at redshifts $z \approx 3$ but
are obscured by dust (Meurer et al.\ 1997), and (3) massive elliptical galaxies
form at higher redshifts, say $z > 6$ (Maoz 1997).

  The possibility that massive elliptical galaxies form at redshifts $z > 6$ is
accessible to direct investigation by means of existing observations of the HDF. 
By combining images obtained at optical wavelengths by HST (Williams et al.\
1996) with images obtained at infrared wavelengths by the Kitt Peak National
Observatory (KPNO) 4 m telescope (Dickinson et al.\ 1998, in preparation),
galaxies may in principle be detected at redshifts as large as $z \approx 17$,
beyond which \lya\ is redshifted past the response of the $K$-band filter.
Such galaxies almost certainly must exhibit the same Lyman absorption signature
characteristic of other high-redshift galaxies---namely complete absorption of
flux below the Lyman limit and strong absorption of flux in the \lya\ forest. 
The reason is that this signature is imprinted by intrinsic and intervening
neutral hydrogen, which is observed to be abundant in the early universe.  The
Lyman absorption signature thus provides a robust and sensitive means of
identifying extremely high redshift galaxy candidates or of establishing an
empirical limit on extremely high redshift galaxies using broad-band photometric
techniques.

  Here we apply the Lyman absorption signature to search for galaxies at
redshifts $z \approx 6 - 17$ using optical and infrared images of the HDF. 
Adopting plausible assumptions to relate rest-frame ultraviolet flux densities
to unobscured star formation rates and adopting standard Friedmann cosmological
models with dimensionless Hubble constant $h = H_0 / (100 \ \kms\ {\rm
Mpc}^{-1})$, deceleration parameter in the range $q_0 = 0 - 0.5$, and zero
cosmological constant, the infrared images are easily sensitive to the star
formation rates expected for massive elliptical galaxy formation to quite high
redshifts.  For $q_0 = 0.5$, the infrared images are sensitive to an unobscured
star formation rate of $\dot{M} = 100 \ h^{-2} \ M_\odot$ yr$^{-1}$ to redshifts
as large as $z = 17$, and for $q_0 = 0$, the infrared images are sensitive to an
unobscured star formation rate of $\dot{M} = 300 \ h^{-2} \ M_\odot$ yr$^{-1}$
to redshifts as large as $z = 14$.  The primary result of the analysis is that
only one extremely high redshift galaxy candidate is identified at the $5
\sigma$ level of significance (and four at the $4 \sigma$ level), which implies
a strict upper limit to the surface density of extremely high redshift galaxies
and a strict upper limit to the volume density of extremely high redshift
galaxies if (and only if) such galaxies are not highly obscured by dust.

\section{METHOD}

  There are two reasons to believe that extremely high redshift galaxies must
exhibit the same Lyman absorption signature that is characteristic of other
high-redshift galaxies---namely complete absorption of flux below the Lyman
limit and strong absorption of flux in the \lya\ forest.  First, extremely high
redshift galaxies are probably optically thick to Lyman continuum absorption by
intrinsic neutral hydrogen.  This is because extremely high redshift galaxies
are probably characterized by much higher gas fractions than nearby galaxies,
which are observed to be optically thick to Lyman continuum absorption by
intrinsic neutral hydrogen (Leitherer et al.\ 1995).  Second, the extremely high
redshift universe is probably optically thick to Lyman continuum and \lya-forest
absorption by intervening neutral hydrogen.  This is because at the highest
redshifts accessible to QSO absorption line measurements ($z \approx 3 - 5$),
the universe is observed to be optically thick to Lyman continuum and
\lya-forest absorption, and the incidence of Lyman continuum and \lya-forest
absorption systems is observed to be increasing rapidly with increasing
redshift.  For example, the \lya-forest flux decrement parameter $D_A$ is
observed to increase monotonically with redshift to a value $D_A = 0.7$ at $z
\approx 5$ (Madau 1995), and a straightforward linear extrapolation of
measurements of $D_A$ over the redshift interval $z \approx 2.5 - 5$ predicts
$D_A = 1$ (i.e.\ complete absorption of flux in the \lya\ forest) by $z \approx
6.2$.  At even higher redshifts---at epochs before galaxy formation depletes the
intergalactic medium and before ionizing photons produced by massive stars
ionize the intergalactic medium---the universe is almost certainly optically
thick to smoothly distributed Lyman continuum and \lya-forest Gunn--Peterson
absorption.

  For these reasons, we conclude that the Lyman absorption signature provides a
robust and sensitive means of identifying extremely high redshift galaxy
candidates or of establishing an empirical limit on extremely high redshift
galaxies using broad-band photometric techniques.  The method is to apply the
Lyman absorption signature to identify extremely high redshift galaxy candidates
using optical and infrared photometry in just the same way that the Lyman
absorption signature is applied to identify other high-redshift galaxy
candidates using optical photometry alone, namely by seeking objects that
exhibit a precipitous spectral discontinuity across \lya.  At redshifts $z > 6$
the Lyman absorption signature is redshifted beyond observed-frame optical
wavelengths, hence extremely high redshift galaxy candidates must be identified
as objects that are (1) detected at infrared wavelengths and (2) not detected to
within sensitive upper limits at optical wavelengths.

  The major uncertainty in using photometric techniques to estimate redshifts of
extremely high redshift galaxies is that owing to the lack of background probes
the optical depth to \lya-forest absorption has not yet been measured at
redshifts beyond $z \approx 5$.  Here we assume that at redshifts $z > 6$ the
optical depth to \lya-forest absorption approaches $\tau \rightarrow \infty$,
which is consistent with the trend established at lower redshifts.  In this
case, at infrared wavelengths galaxies may in principle be detected at redshifts
as large as $z \approx 17$, beyond which \lya\ is redshifted past the response
of the $K$-band filter.

\section{ANALYSIS}

  The optical images were obtained by HST in December, 1995 using the Wide Field
Planetary Camera 2 (WFPC2) and the F300W, F450W, F606W, and F814W filters
(Williams et al.\ 1996).  The infrared images were obtained by the KPNO 4 m
telescope in April, 1996 using the IRIM camera and standard $J$, $H$, and $K$
filters (Dickinson et al.\ 1998, in preparation).  Our previous analysis
(Lanzetta, Yahil, \& Fern\'{a}ndez-Soto 1996) identified objects in the F814W
image, the faintest of which is of total magnitude $AB(8140) \approx 30$.  The
goal of the current analysis is to identify objects in regions of the
infrared images not occupied by objects detected at optical wavelengths.

  First, we measured the point spread functions of the infrared images by
fitting Gaussian profiles to stars detected in the images.  We accounted for
variations of the point spread functions across the images by fitting the
variations of the FWHM of the point spread functions versus positions from the
centers of the images by quadratic polynomials.  Next, we masked objects
detected at optical wavelengths from the infrared images by setting the values
of pixels associated with objects identified in the F814W image equal to zero. 
We located such pixels by modeling the spatial profile of every object detected
in the F814W image as a convolution of the portion of the F814W image containing
the object with the appropriate point spread function of the $K$-band image and
fitting the model spatial profiles to the $K$-band image to define isophotal
mask regions.  Next, we measured the noise characteristics of the infrared
images by determining the empirical covariances of the masked images.  The
infrared images are significantly correlated over adjacent pixels, hence
off-diagonal terms of the empirical covariances must be included in order to
determine accurate photometric uncertainties.  Finally, we searched for objects
detected in the infrared images by forming the ``point source flux images'' and
corresponding ``point source flux variance images'' from the masked images. 
[We expect high-redshift galaxies to appear as point sources in the infrared
images because our previous analysis (Lanzetta, Yahil, \& Fern\'andez-Soto et
al.\ 1996) indicates that high-redshift galaxies are no more than several tenths
arcsec across, which is much less than the ${\rm FWHM} \approx 1$ arcsec point
spread functions of the infrared images.]  The optimally-weighted point source
flux at pixel $i,j$ is
\begin{equation}
F(i,j) = \frac{\Sigma_{i',j'} f(i',j') \Phi(i-i',j-j')}{\Sigma_{i',j'}
\Phi^2(i',j')},
\end{equation}
and the optimally-weighted point source flux variance at pixel $i,j$ is
\begin{equation}
{\rm var}[F(i,j)] = A \frac{\Sigma_{i',j'} \sigma^2(i',j')
\Phi^2(i-i',j-j')}{[\Sigma_{i',j'} \Phi^2(i',j')]^2},
\end{equation}
where $f(i,j)$ is the flux at pixel $i,j$, $\sigma^2(i,j)$ is the flux variance
at pixel $i,j$, $\Phi(i,j)$ is the point spread function at pixel $i,j$, and 
$A$ is a factor of order unity that accounts for off-diagonal terms of the
empirical covariance and where the sums extend over all pixels covered by the
point spread function.
The point source flux
images are shown in Figure 1, and the point source $5 \sigma$ detection
thresholds determined from the point source flux variance images are listed in
Table 1.  Objects detected in the infrared images are identified by comparing
pixels of the point source flux images with corresponding pixels of the point
source variance images.  The limiting magnitude thresholds are $AB(12,000) =
24.3$, $AB(16,000) = 23.7$, and $AB(22,000) = 23.8$.

  The result of the search is that one object at the $5 \sigma$ level of
significance and four objects at the $4 \sigma$ level of significance are
detected in the $K$-band image.  We designate these objects as objects A through
E in order of decreasing flux.  No objects are detected in the $J$- or $H$-band
images at the $5 \sigma$ level of significance.  Properties of the objects are
listed in Table 2, $K$-band and F814W images of the objects are shown in Figure
2, and the spectral energy distributions of the objects are shown in Figure 3. 
The solid angle $\Omega$ of the portion of the infrared images included into the
analysis is $\Omega = 5.1$ arcmin$^2$, and the fraction $f$ of the portion of
the infrared images included into the analysis and associated with objects
identified in the F814W image is $f = 0.33$.  The solid angle $\Omega (1 - f)$
of the portion of the infrared images included into the analysis and not
associated with objects identified in the F814W image is $\Omega (1 - f) = 3.4$
arcmin$^2$.

  We performed several tests of the reality of objects A through E as follows: 
First, we sought to identify negative fluctuations in the images employing
exactly the same techniques used to identify positive fluctuations in the
images.  No negative fluctuations were detected in the $J$-, $H$-, or $K$-band
images at the $4 \sigma$ or greater level of significance.  Next, we estimated
probabilities of detecting chance 4 or $5 \sigma$ fluctuations, assuming a
Gaussian distribution of noise fluctuations.  The portion of the infrared images
included into the analysis is made up of $\approx 4000$  independent resolution
elements, over which the probability of detecting a chance $5 \sigma$
fluctuation is just $1.2 \times 10^{-8}$ and of a chance $4 \sigma$ fluctuation
is just $1.2 \times 10^{-4}$.  Finally, we sought to verify the objects in
independent $K$-band images  obtained by Hogg et al.\ (1996) with the Keck
telescope.  The Keck images (which cover only a small portion of the HDF)
include only object D, which occurs at the very edge of the Keck image 2. 
Object D is present in the Keck image, with a flux consistent with the value
measured from the KPNO image.  We conclude that objects A through E are likely
to be real rather than chance fluctuations.

\section{IMPLICATIONS}

  The primary result of the analysis is that only one extremely high redshift
galaxy candidate is identified at the $5 \sigma$ level of significance (and
four at the $4 \sigma$ level), which implies strict upper limits to the surface
and volume densities of extremely high redshift galaxies.  Here we consider the
$5 \sigma$ level of significance threshold in order to establish conservative
limits that are unlikely to be contaminated by spurious detections.

  Following Steidel et al.\ (1996) and Lowenthal et al.\ (1997), we use the
spectral synthesis models of Leitherer, Robert, \& Heckman (1995) to relate
rest-frame ultraviolet flux densities to unobscured star formation rates.  We
adopt a power-law relationship between monochromatic luminosity density
$L_\lambda$ and wavelength $\lambda$
\begin{equation}
L_\lambda(\lambda) \propto \lambda^{-\beta}
\end{equation}
and express unobscured star formation rate $\dot{M}$ in terms of monochromatic
luminosity density $L_{\lambda_{1500}}$ at rest-frame wavelength $\lambda =
1500$ \AA\ as
\begin{equation}
\dot{M} = \frac{L_{\lambda_{1500}}}{1.4 \times 10^{40} \ {\rm erg} \
{\rm s}^{-1} \ {\rm \AA}^{-1}} \ M_\odot \ {\rm yr}^{-1}.
\end{equation}
Equation (4) applies for a continuous star formation model with a Salpter (1955)
initial mass function of upper mass cutoff $M_{\rm up} = 80 M_\odot$.  A
continuous star formation model is appropriate because the star formation time
scale is expected to be no less than the dynamical time scale, which for massive
elliptical galaxies is much longer than massive star lifetimes.  We adopt $\beta
= 1.2$, which is consistent with measurements of nearby starburst galaxies and
galaxies with spectroscopically determined redshifts $z \approx 3$ in the HDF
(Meurer et al.\ 1997) and with our own measurements of galaxies with
photometrically estimated redshifts $z \approx 5$ in the HDF.  For the
adopted value of $\beta$, the $J$- and $K$-band images reach almost identical
levels of sensitivity to star formation while the $H$-band image reaches a lower
level of sensitivity to star formation.  For this reason, we consider only
the $K$-band image to identify extremely high redshift galaxies at all 
redshifts $z \approx 6 - 17$.

  The unobscured star formation rate corresponding to the $5 \sigma$ detection
threshold of the $K$-band image at redshifts $z = 6 - 17$ is shown in Figure 3.
[For comparison, the unobscured star formation rate corresponding to the $5
\sigma$ detection threshold of the F814W image at redshifts $z = 0 - 6$ is also
shown in Figure 3.  This rate is calculated assuming a limiting magnitude
threshold of $AB(8140) = 28.0$.]  For $q_0 = 0.5$, the infrared images are
sensitive to an unobscured star formation rate of $\dot{M} = 30 \ h^{-2} \
M_\odot$ yr$^{-1}$ to redshifts as large as $z = 12$ and of $\dot{M} = 100 \
h^{-2} \ M_\odot$ yr$^{-1}$ to redshifts as large as $z = 17$.  For $q_0 = 0$,
the infrared images are sensitive to an unobscured star formation rate of
$\dot{M} = 100 \ h^{-2} \ M_\odot$ yr$^{-1}$ to redshifts as large as $z = 7$
and of $\dot{M} = 300 \ h^{-2} \ M_\odot$ yr$^{-1}$ to redshifts as large as $z
= 14$.

  The result that only one extremely high redshift galaxy candidate is
identified at the $5 \sigma$ level of significance establishes a $3 \sigma$
upper limit of 5.2 to the mean number of extremely high redshift galaxies over
the area and volume sampled by the masked images.  The solid angle sampled by
the masked images is 3.4 arcmin$^{-2}$, and the corresponding $3 \sigma$ upper
limit to the surface density $\Sigma$ of extremely high redshift galaxies to a
limiting magnitude threshold $AB(22,000) = 23.8$ is $\Sigma < 1.5$
arcmin$^{-2}$.  The comoving volume sampled by the masked images between
redshifts $z_{\rm min} = 6$ and $z_{\rm max} = 7 - 17$ is listed in Table 3. 
For $q_0 = 0.5$, the corresponding $3 \sigma$ upper limit to the comoving volume
density $n$ of extremely high redshift galaxies with unobscured star formation
rates exceeding $\dot{M} = 30 \ h^{-2} \ M_\odot$ yr$^{-2}$ is $n < 1.8 \times
10^{-3} \ h^3$ Mpc$^3$ and exceeding $\dot{M} = 100 \ h^{-2} \ M_\odot$
yr$^{-2}$ is $n < 1.2 \times 10^{-3} \ h^3$ Mpc$^3$.  For $q_0 = 0.5$, the
corresponding $3 \sigma$ upper limit to the comoving volume density $n$ of
extremely high redshift galaxies with unobscured star formation rates exceeding
$\dot{M} = 100 \ h^{-2} \ M_\odot$ yr$^{-2}$ is $n < 3.7 \times 10^{-4} \ h^3$
Mpc$^3$ and exceeding $\dot{M} = 300 \ h^{-2} \ M_\odot$ yr$^{-2}$ is $n < 3.1
\times 10^{-5} \ h^3$ Mpc$^3$.

\section{DISCUSSION}

  The $3 \sigma$ upper limits to the comoving volume density of extremely high
redshift galaxies derived from analysis of the HDF images are substantially
lower (by factors of between 3 and 180) than the volume density $n = 5.6 \times
10^{-3} \ h^3$ Mpc$^{-3}$ of nearby luminous ($L > L_*$) galaxies (Ellis et al.\
1996).  This implies that either (1) early elliptical galaxies are obscured by
dust, (2) intense star formation in early elliptical galaxies occurs with a
short duty cycle, or (3) massive elliptical galaxies do not form at redshifts
$z > 6$.  Although these possibilities cannot be distinguished on the basis of
the current observations, certain arguments can be brought to bear on their
merits.

  First, it is presumably the case that the very first star formation in early
elliptical galaxies is not obscured by dust, simply because the first dust is
produced by the first generation of stars.  Adopting the plausible assumption
that star formation and the subsequent mixing of dust with gas cannot take
place across an entire galaxy on a time scale that is much less than the
dynamical time scale, then the dust-free phase of early galaxies could last for
a significant fraction of the relevant dynamical time scale, which for massive
elliptical galaxies is $\approx 10^8$ yr.  Unless massive elliptical galaxy
formation occurred at exceptionally high redshifts (i.e. $z > 30$), this phase
of galaxy formation should persist through $z < 17$ and hence be in reach of
direct observation, especially for cosmological models with large values of the
deceleration parameter.  Second, it is unlikely that intense star formation in
early elliptical galaxies occurs on time scales much less than the dynamical
time scale.  Again, unless massive elliptical galaxy formation occurred at
exceptionally high redshifts, this phase of galaxy formation should be in reach
of direct observation.  Although not conclusive, the evidence seems to indicate
that massive elliptical galaxies did not form at redshifts $z > 6$.

  If, in spite of the arguments put forward above, galaxies at the very early
epochs probed by the infrared images of the HDF are obscured by even moderate
amounts of dust, then of course substantially higher star formation rates are
allowed.  For example, recent evidence suggests that galaxies of redshift $z
\approx 3$ suffer extinction of $E(B-V) \approx 0.1$, which corresponds to an
ultraviolet attenuation of a factor of several, depending on details of the
extinction curve (e.g. Pettini et al.\ 1997).  If such a factor applies to the
galaxies probed by the infrared images of the HDF, then star formation rates as
large as $\approx 1000 \ M_\odot$ yr${-1}$ cannot be ruled out by the present
observations.

  The upper limits to the comoving volume density of extremely high redshift
galaxies derived from analysis of the HDF images are far more stringent---both
in terms of sensitivity to star formation and of comoving volume sampled---than
the upper limits derived from other searches for high-redshift galaxies at
infrared wavelengths based on emission lines such as [O II] $\lambda 3727$ and
H$\beta$ (e.g.\ Thompson, Djorgovski, \& Beckwith 1994).  The reason is that the
analysis of the HDF images is based on broad-band photometric techniques, which
probe far greater depths than narrow-band techniques.  The infrared images
obtained by the KPNO 4 m telescope are not the deepest images that can be
obtained with existing telescopes.  We expect that the Lyman absorption
signature applied to even deeper infrared images obtained with ground- or
space-based telescopes will either identify a significant population of
extremely high redshift galaxies or will establish a much more interesting
empirical limit on extremely high redshift galaxies.

  Objects A through E are plausible candidates of galaxies at extremely high
redshifts.  Or they might be very cool stars of surface temperature $T < 1400$
K, although there is some indication that at least some of the objects are
spatially resolved.  These objects are prime targets for future observations.

\acknowledgements

  We are grateful to M.\ Dickinson for providing access to the reduced infrared
images and to an anonymous referee for providing helpful comments.  This
research was supported by NASA grant NAGW--4422 and NSF grant AST--9624216.

\newpage

\begin{center}
\begin{tabular}{p{1.0in}ccc}
\multicolumn{4}{c}{TABLE 1} \\
\multicolumn{4}{c}{POINT SOURCE $5 \sigma$ DETECTION THRESHOLDS} \\
\hline
\hline
& \multicolumn{3}{c}{Detection Threshold} \\
\cline{2-4}
& & $f_\nu$ & $f_\lambda$ \\
\multicolumn{1}{c}{Image} & $AB$ & (erg s$^{-1}$ \cmjj\ Hz$^{-1}$)  & (erg
s$^{-1}$ \cmjj\ \AA$^{-1}$) \\
\hline
$J$ \dotfill & 24.3 & $6.8 \times 10^{-30}$ & $1.4 \times 10^{-19}$ \\
$H$ \dotfill & 23.7 & $1.2 \times 10^{-29}$ & $1.4 \times 10^{-19}$ \\
$K$ \dotfill & 23.8 & $1.1 \times 10^{-29}$ & $6.8 \times 10^{-20}$ \\
\hline
\end{tabular}
\end{center}

\newpage

\setcounter{page}{12}

\begin{center}
\begin{tabular}{p{1.0in}cc}
\multicolumn{3}{c}{TABLE 3} \\
\multicolumn{3}{c}{COMOVING VOLUME} \\
\hline
\hline
& \multicolumn{2}{c}{$V$ (Mpc$^3$)} \\
\cline{2-3}
\multicolumn{1}{c}{$z_{\rm max}$} & $q_0 = 0$ & $q_0 = 0.5$ \\
\hline
 7 \dotfill &    14,020  &    609 \\
 8 \dotfill &    30,037  &   1150 \\
 9 \dotfill &    48,039  &   1633 \\
10 \dotfill &    68,017  &   2069 \\
11 \dotfill &    89,964  &   2464 \\
12 \dotfill &   113,877  &   2825 \\
13 \dotfill &   139,749  &   3157 \\
14 \dotfill &   167,579  &   3462 \\
15 \dotfill &   197,363  &   3745 \\
16 \dotfill &   229,101  &   4008 \\
17 \dotfill &   262,789  &   4253 \\
\hline
\end{tabular}
\end{center}

\newpage

\figcaption{Point source flux images determined from the (a) $J$-, (b) $H$-, and
(c) $K$-band infrared images obtained by the KPNO 4 m telescope.  Angular extent
of each image is $163 \times 163$ arcsec$^2$.  Black represents high flux and
white represents low flux.  Solid gray areas indicate regions of the infrared
images occupied by objects detected at optical wavelengths in the optical images
obtained by HST.  The angular extent of the infrared images is slightly larger
than the angular extent of the optical images, hence no objects are detected at
optical wavelengths along the edges of the infrared images  or above and to the
right of the location of the Planetary Camera (i.e.\ in the upper right
corner).}

\figcaption{$K$-band and F814W images of objects A through E.  Angular extent
of each image is $30 \times 30$ arcsec$^2$.  Black represents high flux and
white represents low flux.  Tic marks indicate objects.  $K$-band images have
been processed for optimal detection of point sources, as described by equation
(1).}

\figcaption{Spectral energy distributions of objects A through E.  Vertical
error bars show flux uncertainties, and horizontal error bars show filter FWHM.}

\figcaption{Logarithm of unobscured star formation rate $\dot{M}$ vs.\ redshift
$z$ for (a) $q_0 = 0.5$ and (b) $q_0 = 0$.  Star formation rate is calculated
according to equations (3) and (4) assuming a limiting magnitude threshold of
$AB(22,000) = 23.8$ at redshifts $z = 6 - 17$ and a limiting magnitude threshold
of $AB(8140) = 28.0$ at redshifts $z = 0 - 6$.}

\end{document}